\documentclass[letterpaper]{article}
\usepackage{aaai}
\usepackage{times}
\usepackage{helvet}
\usepackage{courier}
\usepackage{graphicx}

\long\def\comment#1{}
\newcommand{\commentout}[1]{}

\newcommand{\eqref}[1]{Equation~\ref{#1}}
\newcommand{\figref}[1]{Figure~\ref{#1}}
\newcommand{\tabref}[1]{Table~\ref{#1}}

\begin{document}

\title{Personalizing Image Search Results on Flickr}
\author{Kristina Lerman, Anon Plangprasopchok and  Chio Wong \\
University of Southern California \\
Information Sciences Institute\\
4676 Admiralty Way\\
Marina del Rey, California 90292\\
\{lerman,plangpra,chiowong\}@isi.edu}

\maketitle

\begin{abstract}

The social media site Flickr allows users to upload their
photos, annotate them with tags, submit them to groups, and also to
form social networks by adding other users as contacts.
Flickr offers multiple ways of browsing or searching it. One option
is tag search, which returns all images tagged with a specific
keyword. If the keyword is ambiguous, e.g., ``beetle'' could mean
an insect or a car, tag search results will include many images
that are not relevant to the sense the user had in mind when
executing the query.
We claim that users express their photography interests through
the metadata they add in the form of contacts and image annotations.
We show how to exploit this metadata to personalize search results
for the user, thereby improving search performance. First, we show
that we can significantly improve search precision by
filtering tag search results by user's contacts or a larger social
network that includes those contact's contacts. Secondly, we describe a probabilistic model that takes
advantage of tag information to discover latent topics
contained in the search results. The users' interests can similarly be described
by the tags they used for annotating their images.
The latent topics found by the model are then used to personalize
search results by finding images on topics that are of interest to
the user.

\end{abstract}

\section{Introduction}
The photosharing site Flickr is one of the earliest and more
popular examples of the new generation of Web sites, labeled
\textit{social media}, whose content is primarily user-driven. Other
examples of social media include: blogs (personal online journals
that allow users to share thoughts and receive feedback on them),
Wikipedia (a collectively written and edited online encyclopedia),
and Del.icio.us and Digg (Web sites that allow users to share,
discuss, and rank Web pages, and news stories respectively). The
rise of social media underscores a transformation of the Web as
fundamental as its birth.  Rather than simply searching for, and
passively consuming, information, users are collaboratively
creating, evaluating, and distributing information.  In the near
future, new information-processing applications enabled by social
media will include tools for personalized information discovery,
applications that exploit the ``wisdom of crowds'' (e.g., emergent
semantics and collaborative information evaluation), deeper
analysis of community structure to identify trends and experts,
and many others still difficult to imagine.

Social media sites share four characteristics: (1)~Users create or
contribute content in a variety of media types; (2)~Users annotate
content with tags; (3)~Users evaluate content, either actively by voting
or passively by using content; and (4)~Users create social
networks by designating other users with similar interests as
contacts or friends. In the process of using these sites, users
are adding rich metadata in the form of social networks,
annotations and ratings. Availability of large quantities of this
metadata will lead to the development of new algorithms to solve a
variety of information processing problems, from new
recommendation to improved information discovery algorithms.

In this paper we show how user-added metadata on Flickr can be
used to improve image search results. We claim that users express
their photography interests on Flickr, among other ways, by adding photographers
whose work they admire to their social network and through the
tags they use to annotate their own images. We show how to exploit
this information to personalize search results to the individual user.

The rest of the paper is organized as follows. First, we describe tagging and why it can be viewed
as a useful expression of user's interests, as well as some of the
challenges that arise when working with tags. In
Section ``Anatomy of Flickr'' we describe Flickr and its functionality in
greater details, including its tag search capability. In
Section ``Data collections'' we describe the data sets we have collected from
Flickr, including image search results and user information.
In Sections ``Personalizing by contacts'' and ``Personalizing by tags'' we present
the two approaches to personalize search results for an
individual user by filtering by contacts and filtering by
tags respectively. We evaluate the performance of each method on
our Flickr data sets. We conclude by discussing results and future work.

\section{Tagging for organizing images}
\label{sec:tagging} Tags are keyword-based metadata associated with
some content. Tagging was introduced as a means for users to
organize their own content in order to facilitate searching and
browsing for relevant information. It was popularized by the
social bookmarking site Delicious\footnote{http://del.icio.us},
which allowed users to add descriptive tags to their favorite Web
sites. In recent years, tagging has been adopted by many other
social media sites to enable users to tag blogs (Technorati),
images (Flickr), music (Last.fm), scientific papers (CiteULike),
videos (YouTube), etc.

The distinguishing feature of tagging systems is that they use an
uncontrolled vocabulary. This is in marked contrast to previous
attempts to organize information via formal taxonomies and
classification systems. A formal classification system, e.g.,
Linnaean classification of living things, puts an object in a
unique place within a hierarchy. Thus, a \textbf{tiger}
(\emph{Panthera tigris}) is a carnivorous mammal that belongs to
the genus {\em Panthera}, which also includes large cats, such as
lions and leopards. Tiger is also part of the \emph{felidae}
family, which includes small cats, such as the familiar house cat
of the genus \emph{Felis}.

Tagging is a non-hierarchical and non-exclusive categorization,
meaning that a user can choose to highlight any one of the tagged
object's facets or properties. Adapting the example from Golder
and Huberman~\cite{Golder05}, suppose a user takes an image of a
Siberian tiger. Most likely, the user is not familiar with the
formal name of the species (\emph{P. tigris altaica}) and will tag
it with the keyword ``tiger.'' Depending on his needs or mood, the
user may even tag is with more general or specific terms, such as
``animal,'' ``mammal'' or ``Siberian.'' The user may also note
that the image was taken at the ``zoo'' and that he used his
``telephoto'' lens to get the shot. Rather than forcing the image
into a hierarchy or multiple hierarchies based on the equipment
used to take the photo, the place where the image was taken, type
of animal depicted, or even the animal's provenance, tagging
system allows the user to locate the image by any of its
properties by filtering the entire image set on any of the tags.
Thus, searching on the tag ``tiger'' will return all the images of
tigers the user has taken, including Siberian and Bengal tigers,
while searching on ``Siberian'' will return the images of Siberian
animals, people or artifacts the user has photographed. Filtering
on both ``Siberian'' and ``tiger'' tags will return the
intersection of the images tagged with those keywords, in other
words, the images of Siberian tigers.

As Golder and Huberman point out, tagging systems are vulnerable
to problems that arise when users try to attach semantics to
objects through keywords. These problems are exacerbated in social
media where users may use different tagging conventions, but still
want to take advantage of the others' tagging activities. The
first problem is of homonymy, where the same tag may have
different meanings. For example, the ``tiger'' tag could be
applied to the mammal or to Apple computer's operating system.
Searching on the tag ``tiger'' will return many images unrelated
the carnivorous mammals, requiring the user to sift through
possibly a large amount of irrelevant content. Another problem related
to homonymy is that of polysemy, which arises when a word has
multiple related meanings, such as ``apple'' to mean the company
or any of its products. Another problem is that of synonymy, or
multiple words having the same or related meaning, for example,
``baby'' and ``infant.'' The problem here is that if the user
wants all images of young children in their first year of life,
searching on the tag ``baby'' may not return all relevant images,
since other users may have tagged similar photographs with
``infant.'' Of course, plurals (``tigers'' vs ''tiger'') and many
other tagging idiosyncrasies (''myson'' vs ``son'') may also
confound a tagging system.

Golder and Huberman identify yet another problem that arises when
using tags for categorization --- that of the ``basic level.'' A
given item can be described by terms along a spectrum of
specificity, ranging from specific to general. A Siberian tiger
can be described as a ``tiger,'' but also as a ``mammal'' and
``animal.'' The basic level is the category people choose for an
object when communicating to others about it. Thus, for most
people, the basic level for canines is ``dog,'' not the more
general ``animal'' or the more specific ``beagle.'' However, what
constitutes the basic level varies between individuals, and to a
large extent depends on the degree of expertise. To a dog expert,
the basic level may be the more specific ``beagle'' or ``poodle,''
rather than ``dog.'' The basic level problem arises when different
users choose to describe the item at different levels of
specificity. For example, a dog expert tags an image of a beagle
as ``beagle,'' whereas the average user may tag a similar image as
``dog.'' Unless the user is aware of the basic level variation and
supplies more specific (and more general) keywords during tag
search, he may miss a large number of relevant images.

Despite these problems, tagging is a light weight, flexible
categorization system. The growing number of tagged images
provides evidence that users are adopting tagging on
Flickr~\cite{boyd06}. There is speculation~\cite{Mika05} that collective
tagging will lead to a common informal classification system,
dubbed a ``folksonomy,'' that will be used to
organize all information from all users. Developing value-added
systems on top of tags, e.g., which allow users to better browse
or search for relevant items, will only accelerate wider acceptance of
tagging.

\begin{figure*}[tbhp]
  \center{
  \includegraphics[width=3.5in]{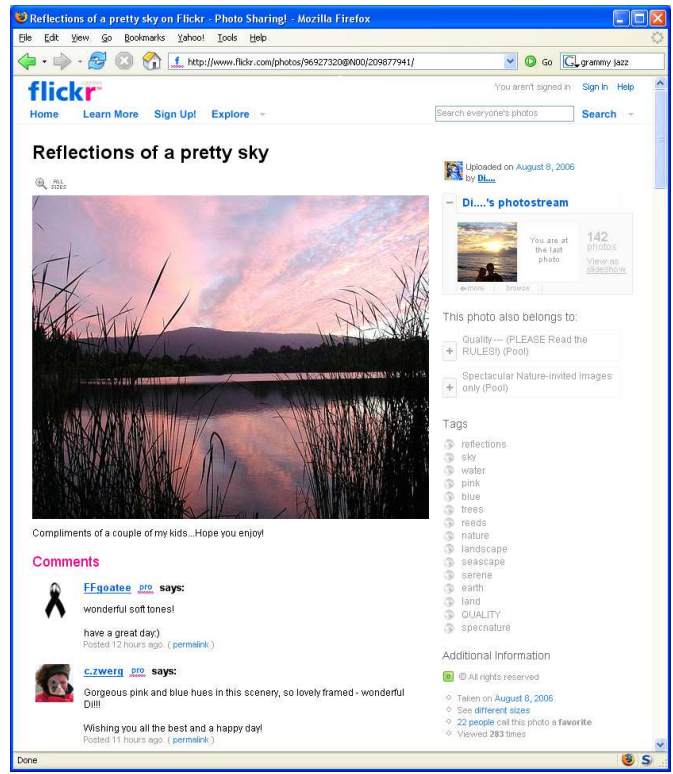}
  }
  \caption{A typical photo page on Flickr}\label{fig:homepage}
  \end{figure*}

\section{Anatomy of Flickr}
\label{sec:flickr} \label{sec:anatomy} Flickr consists of a
collection of interlinked user, photo, tag and group pages. A
typical Flickr photo page is shown in \figref{fig:homepage}. It
provides a variety of information about the image: who uploaded it
and when, what groups it has been submitted to, its tags, who
commented on the image and when, how many times the image was
viewed or bookmarked as a ``favorite.'' Clicking on a user's name
brings up that user's photo stream, which shows the latest photos
she has uploaded, the images she marked as ``favorite,'' and her
profile, which gives information about the user, including a list
of her contacts and groups she belong to. Clicking on the tag
shows user's images that have been tagged with this keyword, or
all public images that have been similarly tagged. Finally, the
group link brings up the group's page, which shows the photo group,
group membership, popular tags, discussions and other information
about the group.

\paragraph{Groups}
Flickr allows users to create special interest groups on any
imaginable topic. There are groups for showcasing exceptional
images, group for images of circles within a square, groups for
closeups of flowers, for the color red (and every other color and
shade), groups for rating submitted images, or those used
solely to generate comments. Some groups are even set up as games,
such as \textsf{The Infinite Flickr}, where the rule is that a user post an
image of herself looking at the screen showing the last image (of
a user looking at the screen showing next to last image, etc).

There is redundancy and duplication in groups. For example, groups
for child photography include \textsf{Children's Portraits},
\textsf{Kidpix}, \textsf{Flickr's Cutest Kids}, \textsf{Kids in
Action}, \textsf{Toddlers}, etc. A user chooses one, or usually
several, groups to which to submit an image. We believe that group
names can be viewed as a kind of publicly agreed upon tags.

\paragraph{Contacts}
Flickr allows users to designate others as friends or
contacts and makes it easy to track their activities. A single
click on the ``Contacts'' hyperlink shows the user the latest images
from his or her contacts. Tracking activities of friends is a
common feature of many social media sites and is one of their
major draws.

\paragraph{Interestingness}
Flickr uses the ``interestingness'' criterion to evaluate the quality
of the image. Although the algorithm that is used to compute this is kept secret to
prevent gaming the system, certain metrics are taken into account:
``where the clickthroughs are coming from; who comments on it and
when; who marks it as a favorite; its tags and many more things
which are constantly
changing.''\footnote{http://flickr.com/explore/interesting/}

\subsection{Browsing and searching}
Flickr offers the user a number of browsing and searching methods.
One can browse by popular tags, through the groups directory,
through the Explore page and the calendar interface, which
provides access to the 500 most ``interesting'' images on any
given day. A user can also browse geotagged images through the
recently introduced map interface. Finally, Flickr allows for
social browsing through the ``Contacts'' interface that shows in one
place the recent images uploaded by the user's designated
contacts.

Flickr allows searching for photos using full text or tag search.
A user can restrict the search to all public photos, his or her own photos,
photos she marked as her favorite, or photos from a specific
contact. The advanced search interface currently allows further filtering by
content type, date and camera.

Search results are by default displayed in reverse chronological
order of being uploaded, with the most recent images on
top. Another available option is to display images by their
``interestingness'' value, with the most ``interesting'' images on
top.

\subsection{Personalizing search results}
Suppose a user is interested in wildlife photography and wants to
see images of tigers on Flickr. The user can search for all public
images tagged with the keyword ``tiger.'' As of March 2007, such a
search returns over $55,500$ results. When images are arranged by
their ``interestingness,'' the first page of results contains many
images of tigers, but also of a tiger shark, cats, butterfly and a
fish. Subsequent pages of search results show, in addition to
tigers, children in striped suits, flowers (tiger lily), more
cats, Mac OS X (tiger) screenshots, golfing pictures (Tiger
Woods), etc. In other words, results include many false positives,
images that are irrelevant to what the user had in mind
when executing the search.

We assume that when the search term is ambiguous, the sense that
the user has in mind is related to her interests. For example,
when a child photographer is searching for pictures of a
``newborn,'' she is most likely interested in photographs of human
babies, not kittens, puppies, or ducklings. Similarly, a nature
photographer specializing in macro photography is likely to be
interested in insects when searching on the keyword ``beetle,''
not a Volkswagen car. Users express their photography preferences
and interests in a number of ways on Flickr. They express them
through their contacts (photographers they choose to watch),
through the images they upload to Flickr, through the tags they
add to these images, through the groups they join, and through the
images of other photographers they mark as their favorite. In this
paper we show that we can personalize results of tag search by
exploiting information about user's preferences. In the sections
below, we describe two search personalization methods: one that
relies on user-created tags and one that exploits user's contacts.
We show that both methods improve search performance by reducing
the number of false positives, or irrelevant results, returned to
the user.

\section{Data collections}
\label{sec:data} To show how user-created metadata can be used to
personalize results of tag search, we retrieved a variety of data from
Flickr using their public API.

\subsection{Data sets}

We collected images by performing a single keyword tag search of all
public images on Flickr. We specified that the returned images are
ordered by their ``interestingness'' value, with most interesting
images first. We retrieved the links to the top 4500 images for
each of the following search terms:
\begin{description}
    \item[tiger] possible senses include (a) big cat ( e.g., Asian
    tiger), (b) shark (Tiger shark), (c) flower (Tiger Lily), (d) golfing
    (Tiger Woods), etc.

    \item[newborn] possible senses include (a) a human baby, (b) kitten,
    (c) puppy, (d) duckling, (e) foal, etc.

    \item[beetle] possible senses include (a) a type of insect and (b) Volkswagen car model

\end{description}

For each image in the set, we used Flickr's API to retrieve
the name of the user who posted the image (image owner), and all
the image's tags and groups.

\subsection{Users}

Our objective is to personalize tag search results; therefore, to
evaluate our approach, we need to have users to whose interests
the search results are being tailored. We identified four users
who are interested in the first sense of each search term.  For
the \emph{newborn} data set, those users were one of the authors of
the paper and three other contacts within that user's social
network who were known to be interested in child photography. For
the other datasets, the users were chosen from among the
photographers whose images were returned by the tag search. We
studied each user's profile to confirm that the user was
interested in that sense of the search term. We specifically
looked at group membership and user's tags. 
Thus, for the \emph{tiger} data set, groups that pointed to the
user's interest in \emph{P. tigris} were \textsf{Big Cats},
\textsf{Zoo}, \textsf{The Wildlife Photography}, etc. In addition
to group membership, tags that pointed to user's interest in a
topic, e.g., for the \emph{beetle} data set, we assumed that users
who used tags \textsf{nature} and \textsf{macro} were interested
in insects rather than cars. Likewise, for the \emph{newborn}
data set, users who had uploaded images they tagged with
\textsf{baby} and \textsf{child} were probably interested in human
newborns.

For each of the twelve users, we collected the names of their
contacts, or \textbf{Level 1 contacts}. For each of these
contacts, we also retrieved the list of their contacts. These are
called \textbf{Level 2 contacts}. In addition to contacts, we also
retrieved the list of all the tags, and their frequencies, that
the users had used to annotate their images. In addition to \textbf{all
tags}, we also extracted a list of \textbf{related tags} for each user. These
are the tags that appear together with the tag used as the search
term in the user's photos. In other words, suppose a user, who is
a child photographer, had used tags such as ``baby'', ``child'',
``newborn'', and ``portrait'' in her own images. Tags related to
\emph{newborn} are all the tags that co-occur with the ``newborn''
tag in the user's own images. This information was also extracted via
Flickr's API.

\subsection{Search results}
\label{sec:plainsearch} We manually evaluated the top 500 images
in each data set and marked each as relevant if it was related to
the first sense of the search term listed above, not relevant or
undecided, if the evaluator could not understand the image well
enough to judge its relevance.

\begin{table}
\begin{tabular}{|r|c|c|c|}
  \hline
  \textbf{query} & \textbf{relevant} & \textbf{not relevant} & \textbf{precision} \\ \hline
  newborn & 412 & 83 & 0.82\\
  tiger & 337 & 156 & 0.67\\
  beetle & 232 & 268 & 0.46\\
  \hline
\end{tabular}
\caption{Relevance results for the top 500 images retrieved by tag search}\label{tbl:rel}
\end{table}

In \tabref{tbl:rel}, we report the precision of the search
within the 500 labeled images, as judged from the point of view of
the searching users. Precision is defined as the ratio of relevant
images within the result set over the 500 retrieved images.
Precision of tag search on these sample queries is not very high
due to the presence of false positives
--- images not relevant to the sense of the search term the user had in mind.
In the sections below we show how to improve
search performance by taking into consideration supplementary
information about user's interests provided by her contacts and tags.

\section{Personalizing by contacts}
\label{sec:contacts} Flickr encourages users to designate others
as contacts by making is easy to view the latest images submitted by
them through the ``Contacts'' interface. Users add
contacts for a variety of reasons, including keeping in touch with
friends and family, as well as to track photographers whose work is of
interest to them. We claim that the latter reason is the most
dominant of the reasons. Therefore, we view user's contacts as
an expression of the user's interests. In this section we show
that we can improve tag search results by filtering through the
user's contacts. To personalize search results for a particular
user, we simply restrict the images returned by the tag search to
those created by the user's contacts.

\begin{table*}
\begin{tabular}{|l|c|c|c|c|c|c||c|c|c|c|c|c|}
  \hline
\small \textbf{user} & \small \textbf{\# L1} & \small\textbf{rel.}
& \small \textbf{not rel.} & \small \textbf{Pr} & \small
\textbf{Re} &\small \textbf{improv} & \small \textbf{\# L2+L2} &
\small \textbf{rel.} & \small \textbf{not rel.} & \small
\textbf{Pr} & \small \textbf{Re} &\small \textbf{improv}
\\ \hline
&\multicolumn{11}{c}{\textbf{newborn}}& \\
\hline

user1 & 719 & 232 & 0  & 1.00 &  0.56 & 22\% & 49,539 & 349 & 62  & 0.85 & 0.85 & 4\%\\
user2 & 154 & 169 & 0  & 1.00 & 0.41 &  22\% & 10,970 & 317 & 37  & 0.9 & 0.77 & 10\%\\
user3 & 174 & 147 & 0  & 1.00 & 0.36 & 22\% & 13,153 & 327 & 39 & 0.89 & 0.79 & 9\%\\
user4 & 128 &  132 & 0 & 1.00 & 0.32 & 22\% & 8,439 &  310 & 29 &
0.91 & 0.75 & 11\%\\
\hline &\multicolumn{11}{c}{\textbf{tiger}} &\\ \hline user5 & 63
&
11 &  1 &  0.92 & 0.03 & 37\% & 13,142 & 255 & 71 & 0.78 & 0.76 & 16\%\\
user6 & 103 &   78 & 3 &  0.96 & 0.23 & 44\% & 14,425 &  266 & 83 & 0.76 & 0.79 & 13\%\\
user7 & 62 & 65 & 1 &  0.98 & 0.19 & 47\% & 7,270 & 226 & 60 & 0.79 & 0.67 & 18\%\\
user8 & 56 &  30 &  0 &   0.97 & 0.09 & 44\% & 7,073 & 240 & 63 &
0.79 & 0.71 & 18\%\\
 \hline
&\multicolumn{11}{c}{\textbf{beetle}} & \\ \hline user9 & 445 &
18
& 1 &  0.95 & 0.08 &  106\% & 53,480 &  215 & 221 & 0.49 & 0.93 & 7\%\\
user10 & 364 &   35 &  8 &  0.81 & 0.15 & 77\% & 41,568 &  208 & 217 & 0.49 & 0.90 & 7\%\\
user11 & 783 &   78 & 25 & 0.75 & 0.34 & 65\% & 62,610 &  218 & 227 & 0.49 & 0.94 & 7\%\\
user12 & 102 &  7  & 1 &  0.88 & 0.03 & 90\% & 14,324 & 163 & 152 & 0.52 & 0.70
 & 13\%\\
  \hline
\end{tabular}
\caption{Results of filtering tag search by user's contacts. ``\#
L1'' denotes the number of Level 1 contacts and ``\# L1+L2'' shows
the number of Level 1 and Level 2 contacts, with the succeeding
columns displaying filtering results of that method: the number of
images marked relevant or not relevant, as well as precision and
recall of the filtering method relative to the top 500 images. The
columns marked ``improv'' show improvement in precision over plain
tag search results.}\label{tbl:socnet}
\end{table*}

\tabref{tbl:socnet} shows how many of the 500 images in each
data set came from a user's contacts. The column labeled ``\# L1''
gives the number of user's Level 1 contacts. The following columns
show how many of the images were marked as relevant  or not
relevant by the filtering method, as well as precision and recall
relative to the 500 images in each data set. Recall measures the
fraction of relevant retrieved images relative to all relevant
images within the data set. The last column ``improv'' shows
percent improvement in precision over the plain (unfiltered)
tag search.

As \tabref{tbl:socnet} shows, filtering by contacts improves the
precision of tag search for most users anywhere from 22\% to over
100\% when compared to plain search results in \tabref{tbl:rel}.
The best performance is attained for users within the
\emph{newborn} set, with a large number of relevant images
correctly identified as being relevant, and no irrelevant images
admitted into the result set. The \emph{tiger} set shows an
average precision gain of 42\% over four users, while the
\emph{beetle} set shows an 85\% gain.

Increase in precision is achieved by reducing the number of false
positives, or irrelevant images that are marked as relevant by the
search method. Unfortunately, this gain comes at the expense of
recall: many relevant images are missed by this filtering method.
In order to increase recall, we enlarge the contacts set by
considering two levels of contacts: user's contacts (Level 1) and
her contacts' contacts (Level 2). The motivation for this is that
if the contact relationship expresses common interests among
users, user's interests will also be similar to those of her
contacts' contacts.

The second half of \tabref{tbl:socnet} shows the performance of
filtering the search results by the combined set of user's Level 1
and Level 2 contacts. This method identifies many more relevant
images, although it also admits more irrelevant images, thereby
decreasing precision. This method still shows precision
improvement over plain search, with precision gain of $9\%$,
$16\%$ and $11\%$ respectively for the three data sets.

\section{Personalizing by tags}
\label{sec:tags} In addition to creating lists of contacts, users
express their photography interests through the images they post
on Flickr. We cannot yet automatically understand the content of
images. Instead, we turn to the metadata added by the user to the
image to provide a description of the image. The metadata comes in
a variety of forms: image title, description, comments left by
other users, tags the image owner added to it, as well as the
groups to which she submitted the image. As we described in the
paper, tags are useful image descriptors, since they are used to
categorize the image. Similarly, group names can be viewed as
public tags that a community of users have agreed on. Submitting
an image to a group is, therefore, equivalent to tagging it with a
public tag.

In the section below we describe a probabilistic model that takes
advantage of the images' tag and group information to discover latent topics
in each search set. The users' interests can similarly be described
by collections of tags they had used to annotate their own images.
The latent topics found by the model can be used to personalize
search results by finding images on topics that are of interest to
a particular user.

\subsection{Model definition}

We need to consider four types of entities in the model: a set of
users $U = \{u_{1},...,u_{n}\}$, a set of images or photos $I = \{i_{1},...,i_{m}\}$,
a set of tags $T = \{t_{1},...,t_{o}\}$, and a set of groups $G =
\{g_{1},...,g_{p}\}$. A photo $i_{x}$ posted by owner
$u_{x}$ is described by a set of tags $\{t_{x1},t_{x2},...\}$ and
submitted to several groups $\{g_{x1},g_{x2},...\}$. The post could be viewed as a tuple
$<i_{x},u_{x},\{t_{x1},t_{x2},...\},\{g_{x1},g_{x2},...\} >$. We
assume that there are $n$ users, $m$ posted photos and $p$ groups
in Flickr. Meanwhile, the vocabulary size of tags is $q$.
In order to filter images retrieved by Flickr in response
to tag search and personalize them for a user $u$, we compute the conditional probability $p(i|u)$,
that describes the probability that the photo $i$ is relevant
to $u$ based on her interests. Images with high enough $p(i|u)$
are then presented to the user as relevant images.

As mentioned earlier, users choose tags from an uncontrolled
vocabulary according to their styles and interests. Images of the
same subject could be tagged with different keywords although they
have similar meaning. Meanwhile, the same keyword could be used to
tag images of different subjects. In addition, a
particular tag frequently used by one user may have a different
meaning to another user. Probabilistic models offer a mechanism for
addressing the issues of synonymy, polysemy and tag
sparseness that arise in tagging systems.

We use a probabilistic topic model~\cite{TopicModelSmyth2004}
to model user's image posting behavior. As in a typical
probabilistic topic model, topics are hidden variables,
representing knowledge categories. In our case, topics are
equivalent to image owner's interests. The process of photo posting
by a particular user could be described as a stochastic process:

\begin{itemize}
\item User $u$ decides to post a photo $i$.

\item Based on user $u$'s interests and the subject of the photo, a set
of topics ${z}$ are chosen.

\item Tag $t$ is then selected based on the set of topics chosen
in the previous state.

\item In case that $u$ decides to expose her photo to some
groups, a group $g$ is then selected according to the chosen topics.

\end{itemize}

\begin{figure}
  \center{
  \includegraphics[width=1.4in]{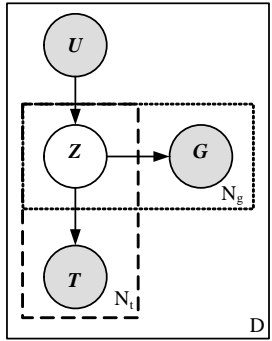}
  }
\caption {Graphical representation for model-based information
filtering. $U$, $T$, $G$ and $Z$ denote variables ``User'',
``Tag'', ``Group'', and  ``Topic'' respectively. $N_{t}$
represents a number of tag occurrences for a one photo (by the
photo owner); $D$ represents a number of all photos on Flickr.
Meanwhile, $N_{g}$ denotes a number of groups for a particular
photo.
 } \label{fig:probmodel}
\end{figure}

The process is depicted in a graphical form in
\figref{fig:probmodel}. We do not treat the image $i$ as a
variable in the model but view it as a co-occurrence of a user, a
set of tags and a set of groups. From the process described above,
we can represent the joint probability of user, tag and group for
a particular photo as

\begin{eqnarray*}
p(i) &=& p(u_{i},T_{i},G_{i})\\
&=& p(u_{i})\cdot\left(\prod_{n_{t}}
\left(\sum_{k}p(z_{k}|u_{i})p(t_{i}|z)\right)^{n_{i}(t)}
\right) \\
&& \cdot
\left(\prod_{n_{g}}\left(\sum_{k}p(z_{k}|u_{i})p(g_{i}|z)
\right)^{n_{i}(g)} \right).
\end{eqnarray*}

Note that it is straightforward to exclude photo's group
information from the above equation simply by omitting the terms relevant
to $g$. $n_{t}$ and $n_{g}$ is a number of all possible tags and
groups respectively in the data set. Meanwhile, $n_{i}(t)$ and
$n_{i}(g)$ act as indicator functions: $n_{i}(t)=1$ if an image
$i$ is tagged with tag $t$; otherwise, it is $0$. Similarly, $n_{i}(g)=1$
if an image $i$ is submitted to group $g$; otherwise, it is 0. $k$ is
the predefined number of topics.

The joint probability of photos in the data set $I$ is defined as
\[
p(I) = \prod_{m}p(i_{m}).
\]
In order to estimate parameters $p(z|u_{i})$, $p(t_{i}|z)$, and
$p(g_{i}|z)$, we define a log likelihood $L$, which measures how the
estimated parameters fit the observed data. According to the EM
algorithm~\cite{EMpaper}, $L$ will be used as an objective function to estimate
all parameters. $L$ is defined as
\[
L(I) = log(p(I)).
\]

In the expectation step (E-step), the joint probability of the hidden
variable $Z$ given all observations is computed from the following
equations:
\begin{equation}
\label{eq:expectedL1} p(z|t,u) \propto p(z|u) \cdot p(t|z)
\end{equation}

\begin{equation}
\label{eq:expectedL2} p(z|g,u) \propto p(z|u) \cdot p(g|z).
\end{equation}

$L$ cannot be maximized easily, since the summation over the
hidden variable $Z$ appears inside the logarithm. We instead
maximize the expected complete data log-likelihood over the hidden
variable, $E[L^{c}]$, which is defined as
\begin{eqnarray*}
E[L^{c}]  &=& \sum_{i}log(p(u))\\
& +& \sum_{i}\sum_{t}n_{i}(t)\cdot\sum_{z}p(z|u,t)\left(log(p(z|u)\cdotp(t|z)\right) \\
& +& \sum_{i}\sum_{g}n_{i}(g)\cdot\sum_{z}p(z|u,g)\left(log(p(z|g)\cdotp(g|z)\right)
\end{eqnarray*}

Since the term $\sum_{i}log(p(u_{i}))$ is not related to
parameters and can be computed directly from the observed data, we
discard this term from the expected complete data log-likelihood.
With normalization constraints on all parameters, Lagrange
multipliers $\tau$, $\rho$, $\psi$ are added to the expected log
likelihood, yielding the following equation
\begin{eqnarray*}
H = E[L^{c}]&+&\sum_{z}\tau_{z}\left(1-\sum_{t}p(t|z)\right)\\
&+&  \sum_{z}\rho_{z}\left(1-\sum_{g}p(g|z)\right)\\
&+&  \sum_{u}\psi_{u}\left(1-\sum_{z}p(z|u)\right).
\end{eqnarray*}

We maximize $H$ with respect to $p(t|z_{k})$, $p(g|z_{k})$, and
$p(z_{k}|u)$, and then eliminate the Lagrange multipliers to
obtain the following equations for the maximization step:
\begin{eqnarray}
\label{mstep1} p(t|z) & \propto & \sum_{m}n_{i}(t) \cdot p(z|t,u)
\\
\label{mstep2} p(g|z) & \propto & \sum_{m}n_{i}(g) \cdot p(z|g,u)
\\
\label{mstep3} p(z_{k}|u_{m}) & \propto & \sum_{m}\big( \sum_{t}
n_{m}(t)\cdot p(z_{k}|u_{m},t)\\
\nonumber & & +\sum_{g} n_{m}(g)\cdot p(z_{k}|u_{m},g)\big)\,.
\end{eqnarray}
The algorithm iterates between E and M step until the log
likelihood for all parameter values converge.

\subsection{Model-based personalization}
We can use the model developed in the previous section to find
the images $i$ most relevant to the interests of a particular user $u^\prime$.
We do so by learning the parameters of the model from the data and
using these parameters to compute the conditional probability
$p(i|u^\prime)$. This probability can be factorized as follows:
\begin{equation}
\label{query1}
p(i|u^\prime) = \sum_z p({u}_{i},T_{i},G_{i}|z) \cdot p(z|u^\prime)\,,
\end{equation}
\noindent where ${u}_{i}$ is the owner of image $i$ in the data
set, and $T_i$ and $G_i$ are, respectively, the set of all the tags and groups for
the image $i$.

The former term in \eqref{query1} can be factorized further as
\begin{eqnarray*}
& p(u_{i},T_{i},G_{i}|z) \propto p(T_{i}|z)\cdotp(G_{i}|z)\cdotp(z|u_{i})\cdot p(u_{i})\\
& =
\left(\prod_{t_{i}}p(t_{i}|z)\right)\cdot\left(\prod_{g_{i}}p(g_{i}|z)\right)\cdot
p(z|u_{i})\cdot p(u_{i})\,.
\end{eqnarray*}
We can use the learned parameters to compute this term
directly.

We represent the interests of user $u^\prime$ as an aggregate of the tags
that $u^\prime$ had used in the past for tagging her own images. This
information is used to to approximate $p(z|u^\prime)$:
\begin{eqnarray*}
    p(z|u^\prime) \propto \sum_{t} n(t^{\prime}=t)\cdot p(z|t)
\end{eqnarray*}
where $n(t^{\prime}=t)$ is a frequency (or weight) of tag
$t^{\prime}$ used by $u^{\prime}$. Here we view $n(t^{\prime}=t)$ is
proportional to $p(t^{\prime}|u^\prime)$. Note that we can use either all
the tags $u^\prime$ had applied to the images in her photostream, or a subset of these tags,
e.g., only those that co-occur with some tag in user's images.

\subsection{Evaluation}
We trained the model separately on each data set of 4500 images.
We fixed the number of topics at ten.
We then evaluated our model-based personalization framework
by using the learned parameters and the information about the interests of the
selected users to compute $p(i|u^\prime)$ for the top 500 (manually
labeled) images in the set. Information about user's interests was
captured either by (1) \emph{all tags} (and their frequencies) that are used in all the images
of the user's photostream or (2) \emph{related tags} that occurred
in images that were tagged with the search keyword (e.g.,
``newborn'') by the user.

Computation of $p(t|z)$ is central to the parameter estimation
process, and it tells us something about how strongly a tag $t$
contributes to a topic $z$. \tabref{tab:ptz} shows the most probable $25$ tags
for each topic for the \emph{tiger} data set trained on ten topics.
Although the tag ``tiger'' dominates most topics, we can discern
different themes from the other tags that appear in each topic.
Thus, topic $z_5$ is obviously about domestic cats, while topic
$z_8$ is about Apple computer products.
Meanwhile, topic $z_ 2$ is about flowers and colors (``flower,'' ``lily,'' ``yellow,'' ``pink,'' ``red'');
topic $z_6$ is about about places (``losangeles,'' ``sandiego,'' ``lasvegas,'' ``stuttgard,''),
presumably because these places have zoos. Topic $z_7$ contains
several variations of tiger's scientific name, ``panthera tigris.''
This method appears to identify related words well. Topic $z_5$, for
example, gives synonyms ``cat,'' ``kitty,'' as well as the more general
term ``pet'' and the more specific terms ``kitten'' and ``tabby.'' It even contains the Spanish version
of the word: ``gatto.'' In future work we plan to explore using this method to
categorize photos in a more abstract way. We also note that related
terms can be used to increase search recall by providing additional
keywords for queries.

\begin{table*}
\begin{tabular}{|c|c|c|c|c|}
\hline \textbf{z}$_1$   &  \textbf{z}$_2$ &   \textbf{z}$_3$ &   \textbf{z}$_4$ &   \textbf{z}$_5$
\\ \hline
tiger   &   tiger   &   tiger   &   tiger   &   tiger   \\
zoo &   specanimal  &   cat &   thailand    &   cat \\
animal  &   animalkingdomelite  &   kitty   &   bengal  &   animal  \\
nature  &   abigfave    &   cute    &   animals &   animals \\
animals &   flower  &   kitten  &   tigers  &   zoo \\
wild    &   butterfly   &   cats    &   canon   &   bigcat  \\
tijger  &   macro   &   orange  &   d50 &   cats    \\
wildlife    &   yellow  &   eyes    &   tigertemple &   tigre   \\
ilovenature &   swallowtail &   pet &   20d &   animalplanet    \\
cub &   lily    &   tabby   &   white   &   tigers  \\
siberiantiger   &   green   &   stripes &   nikon   &   bigcats \\
blijdorp    &   canon   &   whiskers    &   kanchanaburi    &   whitetiger  \\
london  &   insect  &   white   &   detroit &   mammal  \\
australia   &   nature  &   art &   life    &   wildlife    \\
portfolio   &   pink    &   feline  &   michigan    &   colorado    \\
white   &   red &   fur &   detroitzoo  &   stripes \\
dierentuin  &   flowers &   animal  &   eos &   denver  \\
toronto &   orange  &   gatto   &   temple  &   sumatrantiger   \\
stripes &   eastern &   pets    &   park    &   white   \\
amurtiger   &   usa &   black   &   asia    &   feline  \\
nikonstunninggallery    &   impressedbeauty &   paws    &   ball    &   mammals \\
s5600   &   tag2    &   furry   &   marineworld &   sumatran    \\
eyes    &   specnature  &   nose    &   baseball    &   exoticcats  \\
sydney  &   black   &   teeth   &   detroittigers   &   exoticcat   \\
cat &   streetart   &   beautiful   &   wild    &   big \\
\hline\hline \textbf{z}$_6$ & \textbf{z}$_7$ &   \textbf{z}$_8$ &   \textbf{z}$_9$ &  \textbf{z}$_{10}$    \\
\hline
tiger   &   nationalzoo &   tiger   &   tiger   &   tiger   \\
tigers  &   tiger   &   apple   &   india   &   lion    \\
dczoo   &   sumatrantiger   &   mac &   canon   &   dog \\
tigercub    &   zoo &   osx &   wildlife    &   shark   \\
california  &   nikon   &   macintosh   &   impressedbeauty &   nyc \\
lion    &   washingtondc    &   screenshot  &   endangered  &   cat \\
cat &   smithsonian &   macosx  &   safari  &   man \\
cc100   &   washington  &   desktop &   wildanimals &   people  \\
florida &   animals &   imac    &   wild    &   arizona \\
girl    &   cat &   stevejobs   &   tag1    &   rock    \\
wilhelma    &   bigcat  &   dashboard   &   tag3    &   beach   \\
self    &   tigris  &   macbook &   park    &   sand    \\
lasvegas    &   panthera    &   powerbook   &   taggedout   &   sleeping    \\
stuttgart   &   bigcats &   os  &   katze   &   tree    \\
me  &   d70s    &   104 &   nature  &   forest  \\
baby    &   pantheratigrissumatrae  &   canon   &   bravo   &   puppy   \\
tattoo  &   dc  &   x   &   nikon   &   bird    \\
endangered  &   sumatrae    &   ipod    &   asia    &   portrait    \\
illustration    &   animal  &   computer    &   canonrebelxt    &   marwell \\
??  &   2005    &   ibook   &   bandhavgarh &   boy \\
losangeles  &   pantheratigris  &   intel   &   vienna  &   fish    \\
portrait    &   nikond70    &   keyboard    &   schönbrunn  &   panther \\
sandiego    &   d70 &   widget  &   zebra   &   teeth   \\
lazoo   &   2006    &   wallpaper   &   pantheratigris  &   brooklyn    \\
giraffe &   topv111 &   laptop  &   d2x &   bahamas \\
\hline
\end{tabular}
  \caption{Top tags ordered by p(t|z) for the ten topic model of the ``tiger'' data set.}\label{tab:ptz}
\end{table*}

\begin{table*}
  \centering
  \begin{tabular}{|c|c|c||c|c||c|c||c|c||c|c|}
    \hline
    &   \textbf{Pr}  &   \textbf{Re}  &   \textbf{Pr}  &   \textbf{Re}  &   \textbf{Pr}  &   \textbf{Re}  &   \textbf{Pr}  &   \textbf{Re}  &   \textbf{Pr}  &   \textbf{Re}
    \\\hline
        &   \multicolumn{9}{c}{\textbf{newborn}} & \\ \hline
        &   \multicolumn{2}{c}{n=50}&  \multicolumn{2}{c}{n=100}&   \multicolumn{2}{c}{n=200} & \multicolumn{2}{c}{n=300}   &\multicolumn{2}{c}{n=412*}\\\hline
user1   &   1.00    &   0.12    &   1.00    &   0.24    &   1.00    &   0.49    &   0.94    &   0.68    &   0.89    &   0.89    \\
user2   &   1.00    &   0.12    &   1.00    &   0.24    &   1.00    &   0.49    &   0.92    &   0.67    &   0.87    &   0.87    \\
user3   &   1.00    &   0.12    &   0.88    &   0.21    &   0.84    &   0.41    &   0.85    &   0.62    &   0.89    &   0.89    \\
user4   &   1.00    &   0.12    &   0.99    &   0.24    &   1.00    &   0.48    &   0.94    &   0.69    &   0.89    &   0.89    \\
    \hline
        &   \multicolumn{9}{c}{\textbf{tiger}} & \\ \hline
    &   \multicolumn{2}{c}{n=50}&  \multicolumn{2}{c}{n=100}&
\multicolumn{2}{c}{n=200} & \multicolumn{2}{c}{n=300}
&\multicolumn{2}{c}{n=337*}\\\hline
user5   &   0.94    &   0.14    &   0.90    &   0.27    &   0.82    &   0.48    &   0.80    &   0.71    &   0.79    &   0.79    \\
user6   &   0.76    &   0.11    &   0.80    &   0.24    &   0.79    &   0.47    &   0.77    &   0.69    &   0.77    &   0.77    \\
user7   &   0.94    &   0.14    &   0.90    &   0.27    &   0.82    &   0.48    &   0.80    &   0.71    &   0.79    &   0.79    \\
user8   &   0.90    &   0.13    &   0.88    &   0.26    &   0.82    &   0.49    &   0.79    &   0.71    &   0.79    &   0.79    \\
    \hline
        &   \multicolumn{9}{c}{\textbf{beetle}} & \\ \hline
        &   \multicolumn{2}{c}{n=50}&  \multicolumn{2}{c}{n=100}&   \multicolumn{2}{c}{n=200} & \multicolumn{2}{c}{n=232*}   &\multicolumn{2}{c}{n=300}\\\hline
user9   &   1.00    &   0.22    &   0.99    &   0.43    &   0.77    &   0.66    &   0.70    &   0.70    &   0.66    &   0.85    \\
user10  &   0.98    &   0.21    &   0.99    &   0.43    &   0.77    &   0.66    &   0.70    &   0.70    &   0.66    &   0.85    \\
user11  &   0.98    &   0.21    &   0.93    &   0.40    &   0.50    &   0.43    &   0.51    &   0.51    &   0.50    &   0.65    \\
user12  &   1.00    &   0.22    &   0.99    &   0.43    &   0.77    &   0.66    &   0.70    &   0.70    &   0.66    &   0.85    \\
    \hline
  \end{tabular}
  \caption{Filtering results where a number of learned topics is 10, excluding group information, and
  user's personal information obtained from all tags she used for her photos. Asterisk denotes R-precision of the
  method, or precision of the first $n$ results, where $n$ is the
  number of relevant results in the data set. }\label{tbl:alltags}
\end{table*}

\tabref{tbl:alltags} presents results of model-based personalization
for the case that uses information from all of user's tags.
The model was trained with ten topics. Results are presented
for different thresholds. The first two columns, for example, report
precision and recall for a high threshold that marks only the 50
most probable images as relevant. The remaining 450 images are
marked as not relevant to the user. Recall is low, because many
relevant images are excluded from the results for such a high
threshold. As the threshold is decreased ($n=100$, $n=200$,
$\ldots$), recall relative to the 500 labeled images increases.
Precision remains high in all cases, and higher than precision of
the plain tag search reported in \tabref{tbl:rel}. In fact, most of the images in the top 100
results presented to the user are relevant to her query. The column
marked with the asterisk gives the R-precision of the method, or
precision of the first $R$ results, where $R$ is the number of
relevant results. The average R-precision of this filtering method
is $8\%$, $17\%$ and $42\%$ better than plain search precision on our three data sets.

\begin{table*}
  \centering
  \begin{tabular}{|c|c|c|c|c|c|c|c|c|c|c|}
    \hline
    &   \textbf{Pr}  &   \textbf{Re}  &   \textbf{Pr}  &   \textbf{Re}  &   \textbf{Pr}  &   \textbf{Re}  &   \textbf{Pr}  &   \textbf{Re}  &   \textbf{Pr}  &   \textbf{Re}
    \\\hline
        &   \multicolumn{9}{c}{\textbf{newborn}} & \\ \hline
        &   \multicolumn{2}{c}{n=50}&  \multicolumn{2}{c}{n=100}&   \multicolumn{2}{c}{n=200} & \multicolumn{2}{c}{n=300}   &\multicolumn{2}{c}{n=412*}\\\hline
    &   Pr  &   Re  &   Pr  &   Re  &   Pr  &   Re  &   Pr  &   Re  &   Pr  &   Re
    \\\hline
user1   &   0.8 &   0.10    &   0.78    &   0.19    &   0.79    &   0.38    &   0.77    &   0.56    &   0.79    &   0.79    \\
user2   &   0.8 &   0.10    &   0.82    &   0.20    &   0.80    &   0.39    &   0.77    &   0.56     &   0.83    &   0.83    \\
user3   &   0.98    &   0.12    &   0.88    &   0.21    &   0.84    &   0.41    &   0.80    &   0.58    &   0.85    &   0.85    \\
user4   &   0.98    &   0.12    &   0.88    &   0.21    &   0.84    &   0.41    &   0.85    &   0.62    &   0.88    &   0.88    \\
    \hline
        &   \multicolumn{9}{c}{\textbf{tiger}} & \\ \hline
    &   \multicolumn{2}{c}{n=50}&  \multicolumn{2}{c}{n=100}&
\multicolumn{2}{c}{n=200} & \multicolumn{2}{c}{n=300}
&\multicolumn{2}{c}{n=337*}\\\hline
user5   &   0.84    &   0.12     &   0.86    &   0.26    &   0.78    &   0.46    &   0.78    &   0.69    &   0.77    &   0.77    \\
user6   &   0.72    &   0.11    &   0.79    &   0.23    &   0.78     &   0.46    &   0.76    &   0.68    &   0.76    &   0.76    \\
user7   &   0.72    &   0.11    &   0.78    &   0.23    &   0.78    &   0.46    &   0.76    &   0.68     &   0.76    &   0.76    \\
user8   &   0.9 &   0.13    &   0.82    &   0.24    &   0.80    &   0.47    &   0.78    &   0.69    &   0.78    &   0.78    \\
    \hline
        &   \multicolumn{9}{c}{\textbf{beetle}} & \\ \hline
        &   \multicolumn{2}{c}{n=50}&  \multicolumn{2}{c}{n=100}&   \multicolumn{2}{c}{n=200} & \multicolumn{2}{c}{n=232*}   &\multicolumn{2}{c}{n=300}\\\hline
user9   &   0.78    &   0.17    &   0.62    &   0.27    &   0.58    &   0.50    &   0.54    &   0.54    &   0.53    &   0.68    \\
user10  &   0.98    &   0.21    &   0.88    &   0.38    &   0.77    &   0.66    &   0.72    &   0.72    &   0.65    &   0.84    \\
user11  &   0.96    &   0.21    &   0.74    &   0.32    &   0.62    &   0.53    &   0.59    &   0.59    &   0.56    &   0.72    \\
user12  &   0.98    &   0.21    &   0.99    &   0.43    &   0.77    &   0.66    &   0.70    &   0.70    &   0.66    &   0.85    \\
    \hline
  \end{tabular}
  \caption{Filtering results where a number of learned topics is 10, excluding group information, and
  user's personal information obtained from all tags she used for her photos, which are tagged by the search term}\label{tbl:relatedtags}
\end{table*}

Performance results of the approach that uses related tags instead
of all tags are given in \tabref{tbl:relatedtags}. We explored this
direction, because we believed it could help discriminate between
different topics that interest a user. Suppose, a child photographer
is interested in nature photography as well as child portraiture.
The subset of tags he used for tagging his ``newborn'' portraits will be
different from the tags used for tagging nature images. These tags
could be used to differentiate between newborn baby and newborn colt
images. However, on the set of users selected for our study, using related
tags did not appear to improve results.
This could be because the tags a particular user used together with, for example,
``beetle'' do not overlap significantly with the rest of the data set.

Including group information did not significantly improve results (not presented in this
manuscript). In fact, group information sometimes hurts the estimation rather than helps.
We believe that this is because our data sets (sorted by Flickr according to image interestingness)
are biased by the presence of general topic groups (e.g., \textsf{Search the Best}, \textsf{Spectacular Nature},
\textsf{Let's Play Tag}, etc.). We postulate that group information would help estimate $p(i|z)$ in cases
where the photo has few or no tags. Group information would help filling
in the missing data by using group name as another tag.
We also trained the model on the data with 15 topics, but found no
significant difference in results.

\section{Previous research}
Recommendation or personalization systems can be categorized into
two main categories. One is collaborative filtering~\cite{Breese98}
which exploits item ratings from many users
to recommend items to other like-minded users. The other is
content-based recommendation, which relies on the contents of an item and
user's query, or other user information, for prediction~\cite{mooney00contentbased}.
Our first approach, filtering by contacts, can be
viewed as implicit collaborative filtering, where the user--contact
relationship is viewed as a preference indicator: it assumes that
the user likes all photos produced by her contacts. In our previous work,
we showed that users do indeed agree with the recommendations made by contacts~\cite{Lerman07digg,Lerman07flickr}.
This is similar to the ideas implemented by MovieTrust~\cite{Golbeck06},
but unlike that system, social media sites do not require users to rate their trust
in the contact.

Meanwhile, our second approach, filtering by tags (and groups),
shares some characteristics with both methods. It is similar to
collaborative filtering, since we use tags to represent agreement between users.
It is also similar to content-based recommendation, because we represent image content
by the tags and group names that have been assigned to it by the user.

Our model-based filtering system is technically similar to, but conceptually different from,
probabilistic models proposed
by \cite{Popescul01}. Both models are probabilistic generative models that
describe co-occurrences of users and items of interest. In particular, the model
assumes a user generates her topics of interest; then the topics generate documents
and words in those documents if the user prefers those documents. In our model, we
metaphorically assume the photo owner generates her topics of interest. The topics,
in turn, generate tags that the owner used to annotate her photo. However, unlike the previous work,
we do not treat photos as variables, as they do for documents. This is because images are tagged only by their
owners;
meanwhile, in their model, all users who are interested in a document generate topics for that document.

Our model-based approach is almost identical to the author-topic model\cite{TopicModelSmyth2004}. However, we extend their
framework to address (1) how to exploit photo's group information for personalized information filtering;
(2) how to approximate user's topics of interest from partially observed personal information (the tags
the user used to describe her own images). For simplicity, we use the classical EM algorithm to train the model;
meanwhile they use a stochastic approximation approach
due to the difficulty involved in performing exact an inference for their generative model.

\section{Conclusions and future work}
\label{sec:conclusion}
We presented two methods for personalizing results of image search
on Flickr. Both methods rely on the metadata users create through
their everyday activities on Flickr, namely user's contacts and the
tags they used for annotating their images.
We claim that this information captures user's tastes and
preferences in photography and  can be used to
personalize search results to the individual user. We showed
that both methods dramatically increase search precision.
We believe that increasing precision is an important goal
for personalization, because dealing with the information overload is
the main issue facing users, and we can help users by reducing the number of
irrelevant results the user has to examine (false positives). Having
said that, our tag-based approach can also be used to expand the
search by suggesting relevant related keywords (e.g.,
``pantheratigris,'' ``bigcat'' and ''cub'' for the query
\emph{tiger}).

In addition to tags and contacts, there exists other metadata,
favorites and comments, that can be used to aid information
personalization and discovery. In our future work we plan to address
the challenge of combing these heterogeneous sources of evidence
within a single approach. We will begin by combining contacts
information with tags.

The probabilistic model needs to be explored further. Right now,
there is no principled way to pick the number of latent topics that are
contained in a data set. We also plan to have a better mechanism for
dealing with uninformative tags and groups. We would like to
automatically identify general interest groups, such as the \textsf{Let's
Play Tag} group, that do not help to discriminate between topics.

The approaches described here can be applied to other social media
sites, such as Del.icio.us. We imagine that in near future, all of
Web will be rich with metadata, of the sort described here, that
will be used to personalize information search and discovery to the
individual user.

\section*{Acknowledgements} This research is based on work supported
in part by the National Science Foundation under Award Nos.
IIS-0535182 and in part by DARPA under Contract No. NBCHD030010.

The U.S.Government is authorized to reproduce and distribute reports
for Governmental purposes notwithstanding any copyright annotation
thereon. The views and conclusions contained herein are those of the
authors and should not be interpreted as necessarily representing
the official policies or endorsements, either expressed or implied,
of any of the above organizations or any person connected with them.

\bibliographystyle{named}
\bibliography{../social,../lerman,../robots,anon}

\end{document}